\documentclass[10pt]{article}
\usepackage{mathrsfs}
\usepackage{amsmath}
\usepackage{amssymb}
\usepackage{cite}
\usepackage{graphicx}
\usepackage{color}
\usepackage{cite}
\usepackage{algorithm}
\usepackage{algorithmic}
\usepackage{multirow}
\usepackage{indentfirst}

\def\qed{ \rule{.1in}{.1in}}
\def\eq#1{\begin{equation}#1\end{equation}}

\newcommand{\matt}[1]{\left[\begin{matrix} #1 \end{matrix}\right]}

\newcommand{\sgn}{{\rm sgn\;}}
\newcommand{\image}{{\rm image\;}}
\newcommand{\col}{{\rm col\;}}
\newcommand{\diag}{{\rm diag\;}}
\newtheorem{theorem}{Theorem}
\newtheorem{assumption}{Assumption}
\newtheorem{lemma}{Lemma}

\newtheorem{remark}{Remark}

\newtheorem{proposition}{Proposition}
\begin{document}

\title{Finite-Time Distributed Linear Equation Solver for Minimum $l_1$ Norm Solutions}
\author{Jingqiu Zhou, Wang Xuan, Shaoshuai Mou, and Brian. D. O. Anderson
\thanks{This work was supported by a funding from Northrop Grumman Cooperation. J. Zhou, X. Wang and S. Mou are with the School of Aeronautics and Astronautics, Purdue University, West Lafayette, IN 47906 USA (e-mail: wang3156@purdue.edu, zhou745@purdue.edu, mous@purdue.edu). B. D. O. Anderson is with Hangzhou Dianzi University, Hangzhou, China,  The Australian National University and Data-61 CSIRO (formerly NICTA), Canberra, ACT 2600 Australia, {e-mail: Brian.Anderson@anu.edu.au}; his work is supported by Data-61, CSIRO, and by the Australian Research Council's Discovery Projects  DP-130103610 and DP-160104500. Corresponding Author: Shaoshuai Mou.}}
\maketitle
\begin{abstract}
   This paper proposes distributed algorithms for multi-agent networks to achieve a solution in finite time to a linear equation $Ax=b$ where $A$ has full row rank, and with the minimum $l_1$-norm in the underdetermined case (where $A$ has more columns than rows). The underlying network is assumed to be undirected and fixed, and an analytical proof is provided for the proposed algorithm to drive all agents' individual states to converge to a common value, viz a solution of $Ax=b$, which is the minimum $l_1$-norm solution in the underdetermined case. Numerical simulations are also provided as validation of the proposed algorithms.
\end{abstract}

\section{Introduction}
	A significant amount of effort in the control community has recently been given to distributed algorithms for solving linear equations over multi-agent networks, in which each agent only knows part of the equation and controls a state vector that can be looked at as an estimate of the solution of the overall linear equations \cite{SJA15TAC,BSUA16NACO,JC16CNS,JN16ACC,SA13ECC}. Numerous extensions along this direction include achieving solutions with the minimum Euclidean norm \cite{XSD17TIE,PWZ16CDC}, elimination of the initialization step \cite{LDA16ACC}, reduction of state vector dimension by utilizing the sparsity of the linear equation \cite{SZLDA16SCL} and achieving least square solutions \cite{JN12ACC,JN11ECC,BJ14TAC,GB16ACC,LYLCABDO2017IFAC,LYLCABDO2017IEEE}. All these algorithms yield asymptotic convergence, but require an infinite number of sensing or communication events.

Solutions to underdetermined linear equations with the minimum $l_1$ norm are perhaps the most important in many engineering applications including earthquake location detection \cite{PMSR97} , analysis of statistical data \cite{Dodge12}, solving biomeganetic inverse problems \cite{RBHAS96}, and so on. One most intriguing case among these applications is compressive sensing, which enables transmission of sparse data in a very efficient way \cite{DDMFWMBSSB09}. The decoding process of compressive sensing requires solving of linear equations with a minimum number of non-zero entries of the solution vectors, which, however, is an NP-hard problem and usually computationally costly \cite{YG12book}. Thus researchers usually turn to achieve solutions with minimum $l_1$ norm instead for which the function to be minimized is convex \cite{ET05TIT,EJT06TIT}. Most existing results for achieving minimum $l_1$ norm solutions are based on the idea of Lasso \cite{AYYAG10} including Alternating Direction Method of Multipliers (ADMM) \cite{SN11ADMM},  the Primal-Dual Interior-Point Method \cite{FKR55TLPMCP, KMMNMS93TCLP}, Gradient Projection Methods  \cite{FMNRD07GPSR}, Homotopy Methods \cite{OMRPBTB00AP}, Iterative Shrinkage-Thresholding Methods \cite{IDMDCD04} and Proximal Gradient Methods\cite{SBJBEJC11}. Interesting as these results are, they either achieve limited accuracy dominated by a threshold parameter, involve solving a much larger linear equation, or lead to high computational complexity.

In this paper we aim to develop distributed algorithms for multi-agent networks to achieve in \textbf{finite time} a solution of linear equations or, in the under-determined case, one with the minimum $l_1$ norm. By distributed is meant that each agent only knows part of the overall linear equation and can communicate with only its nearby neighbors. The problem of interest is formulated in Section \ref{sect2}. We introduce in section \ref{sect3} the concepts to be employed in the paper including  Filippov set-valued maps, Filippov solutions, generalized Lie derivatives, based on which a preliminary result is achieved. In Section \ref{sect4}, we will first propose a distributed algorithm to drive all agents' state vectors to converge in finite time to the same solution of the overall linear equations. Then we present a centralized update for achieving a solution with the minimum $l_1$ norm. Motivated by the projection-consensus flow proposed in \cite{GB16ACC} and the finite-time gradient flow for consensus devised in \cite{CAO08AGREEING,CAO08REACHING,MCAO05LOWERBOUND,SUN2016CONCENSUS,QIN14EXPONENTIAL,J2006Auto}, we utilize a combination of the proposed distributed linear equation solver and the proposed centralized algorithm for minimum $l_1$-norm solutions to develop a distributed linear equation solver for achieving the minimum $l_1$ norm solution, which is shown to converge in finite time. We provide simulations in Section \ref{sect5} and concluding remarks in Section \ref{Sec_con}.
\\

\noindent{ \emph{Notation}:} Let $r$ denote an arbitrary positive integer. Let ${\bf 1}_r$ denote the vector in $\mathbb{R}^r$ with all entries equal to 1s. Let $I_r$ denote the $r\times r$ identity matrix.  We let $\col\{A_1,A_2,\cdots,A_r\}$ be a stack of matrices $A_i$ possessing the same number of columns with the index in a top-down ascending order, $i=1,2,\cdots,r$.
	Let  $\diag\{A_1,A_2,\cdots,A_r\}$ denote a block diagonal matrix with $A_i$ the $i$th diagonal block entry, $i=1,2,\cdots,r$.
	%Let  $(\cdot)_{ij}$ be the ${ij}$th entry of a matrix; $[\cdot]_{ij}$ be the ${ij}$th block of a block matrix.
	By $M> 0$ and $M\geq 0$ are meant that the square matrix $M$ is positive definite and positive semi-definite, respectively. By $M'$ is meant the transpose of a matrix $M$. Let $\ker M$ and $\image M$ denote the kernel  and image of a matrix $M$, respectively. Let $\otimes$ denote the Kronecker product. Let $\|\cdot\|_1$ denote the $l_1$ norm of a vector in $\mathbb {R}^r$.
\section{Problem Formulation}\label{sect2}
Consider a network of $m$ agents, $i=1,2,...,m$; inside this network, each agent can observe states of  certain other agents called its \emph{neighbors}. Let  $\mathcal{N}_i$ denote the set of agent $i$'s neighbors.
	We assume that the neighbor relation is symmetric, that is, $j\in \mathcal{N}_i$ if and only if $i\in \mathcal{N}_j$. Then all these neighbor relations can be described by an $m$-node-$\bar{m}$-edge undirected graph $\mathbb{G}$ such that there is an undirected edge connecting $i$ and $j$ if and only if $i$ and $j$ are neighbors. In this paper we only consider the case in which $\mathbb{G}$ is connected, fixed and undirected.
	
Suppose that each agent $i$ knows $A_i\in \mathbb{R}^{n_i\times n}$ and $b_i\in \mathbb{R}^{n_i}$ and controls a state vector $y_i(t)\in \mathbb{R}^n$. Then all these $A_i$ and $b_i$ can be stacked into an overall equation $Ax=b$, where $A=\col \{A_1,A_2,\cdots\!,A_m\}$, $b=\col \{b_1,b_2,\cdots\!,b_m\}$. Without loss of generality for the problems of interest to us, we assume $A$ to have full row-rank. Let $x^*$ denote a solution to $Ax=b$ (and in the underdetermined case, $x^*$ is not unique); and let $\bar{x}^*$ denote its \emph{ minimum $l_1$-norm solution}, that is,\eq{\bar{x}^*=\arg\min_{Ax=b}  \|x\|_{1} \label{Problem0}} (In the non-singular $A$ case, $x^*$ and $\bar x^*$ necessarily coincide) The problem of interest in this paper is to develop distributed algorithms for each agent $i$ to update its state vector $y_i(t)$ by only using its neighbors' states such that all $y_i(t)$ to converge in finite time to a common value $x^*$ and if desired in the nonsquare case the value $\bar{x}^*$.

\section{Key Concepts and Preliminary Results}{\label{sect3}}
Before proceeding, we introduce some key concepts and preliminary results for future derivation and analysis. The key references for the background we summarize are \cite{Cortes08CSM} and \cite{Filippovbook}.

\subsection{ Filippov set-valued maps and Filippov Solutions}

By a \emph{Filippov set-valued map} $F[f]: \mathbb{R}^r \to\mathcal{B}\subset \mathbb{R}^r$ associated with a function $f : \mathbb{R}^r \to  \mathbb{R}^r$ is meant
 \begin{equation}
 F[f](x) \triangleq \bigcap\limits_{\delta  > 0}\bigcap\limits_{\mu (\mathcal{S})=0}  {\overline {\rm co} } \{ f(B(x,\delta ))/ \mathcal{S}\}
 \label{filiset}
 \end{equation}
 Here $B(x,\delta)$ stands for the open ball on $\mathbb{R}^r$, whose center is at $x$ and has a radius of $\delta$; $\mu(\mathcal{S})$ denotes the Lebesgue measure of $\mathcal{S}$; and $\overline { \rm{co}}$ stands for the convex closure. Let $\sgn(x):\mathbb{R}^r\rightarrow \mathbb{R}^r$ be a function with the $k$th entry, $k=1,2,...,r$, defined as \eq{(\sgn(x))_k=\left\{
 \begin{array}{ll}
  1, & \hbox{$(x)_k>0$;} \\
   -1, & \hbox{$(x)_k<0$;} \\
   0, & \hbox{$(x)_k=0$.}
   \end{array}
   \right. } It follows that the Filippov set-valued map $F[\sgn](x)$ for $x\in \mathbb{R}^r$ is defined entrywise as: \eq{(F[\sgn](x))_k = \left\{ {\begin{array}{*{20}{c}}
{\begin{array}{*{20}{c}}
 					1&{(x)_k > 0}
 				\end{array}} \\
 				{\begin{array}{*{20}{c}}
 						{[ - 1,1]}&{(x)_k = 0}
 					\end{array}} \\
 					{\begin{array}{*{20}{c}}
 							{ - 1}&{(x)_k < 0}
 						\end{array}}
 					\end{array}} \right.\label{eq_flip}} for $k=1,2,...,r$. Note that even if $(x)_i=(x)_j=0$, the $i$th and $j$th entries of a vector in $F[\sgn](x)$ may not necessarily be equal since each of them could be chosen as arbitrary values in the interval $[-1,1]$. From the definition of $F[\sgn](x)$, one can verify that \eq{\label{eq_xl1}q'x=\|x\|_1,\quad \forall q\in F[\sgn](x)}
While for any $w\in\mathbb {R} ^r$, there holds \eq{\label{ineq}q'w\leq\|w\|_1}.

By \emph{a Filippov solution} for  $\dot x  \in F[f](x)$ is meant a Caratheodory solution $x(t)$ such that $\dot x \in F[f](x)$ for almost all $t$, $x(t)$ is absolutely continuous and can be written in the form of an indefinite integral. The following two lemmas treat existence of such a Filippov solution.
\begin{lemma} \label{Lemma_Fili} ( Proposition 3 in \cite{Cortes08CSM})
	If $f : \mathbb{R}^r \to  \mathbb{R}^r$ is measurable and locally bounded, then for any initial point $x_0\in\mathbb{R}^r$, there exists a Filippov solution\footnote{There is no implication that the solution exists on an infinite interval} for  $\dot{x} \in F[f] (x)$.
\end{lemma}
\smallskip
		\begin{lemma}\label{genfilip} ( Theorem $8$ in page $85$ of \cite{Filippovbook} ) Let a vector-valued $f(t,x)$ be defined almost-everywhere in the domain $G$ of time space $(t,x)$. With $f(t,x)$ measurable and locally bounded almost-everywhere in an open domain $G$, let 	\begin{equation}
			F[f](t,x) \triangleq \bigcap\limits_{\delta  > 0}\bigcap\limits_{\mu (\mathcal{S})=0}  {\overline {\rm co} } \{ f(t,B(x,\delta)/\mathcal{S})\}.
			\end{equation}Then for any point $(t_0, x_0)\subset G$, there exists a Filippov solution of $\dot x\in F[f](t,x)$ with $x(t_0) = x_0$.
\end{lemma} 	
\smallskip

Note that Lemma \ref{genfilip} establishes the existence of a solution for time-varying systems. This is more general than Lemma \ref{Lemma_Fili}, which only guarantees the existence of solutions to time-invariant systems.

\subsection{ Generalized Gradients and Generalized Lie Derivatives} For a locally Lipschitz function $w : \mathbb{R}^r \to  \mathbb{R}$, the
\emph{generalized gradient} of $w$  is
\begin{equation}
\partial w(x) \triangleq {\rm co} \{ \mathop {\lim \limits_{i \to \infty } \nabla w({x_i})} :{x_i} \to x,{x_i} \notin \mathcal{S}\cup {\Omega _w}\}
\label{grd}
\end{equation} where $\mathcal{S}\subset \mathbb{R}^r$ is an arbitrarily chosen set of measure zero, ${\Omega}_{w}$ denotes the set of points at which $w$ is not differentiable, and $\rm co$ denotes convex hull. Specially, for the function $||x||_1$, one computes the $k$th element of its generalized gradient to be: \eq{\label{eq_partialx}(\partial ||x||_1)_k = \left\{ {\begin{array}{*{20}{c}}
	{\begin{array}{*{20}{c}}
		1&{x_k > 0}
		\end{array}} \\
	{\begin{array}{*{20}{c}}
		{[ - 1,1]}&{x_k = 0}
		\end{array}} \\
	{\begin{array}{*{20}{c}}
		{ - 1}&{x_k < 0}
		\end{array}}
	\end{array}} \right.}
It follows from this and the definition of $ F[\sgn](x)$ in (\ref{eq_flip}) that
\begin{equation}
F[\sgn](x)=\partial{\|x\|_1}
\label{equa}
\end{equation}

For a set-valued map $\mathcal F:\mathbb R^r\rightarrow\mathcal B(\mathbb R^r)$, the \emph{generalized Lie derivative} of $w$ is defined as
\begin{align}
&\tilde{\mathcal{L}}_{\mathcal{F}}w(x) = \{ q \in \mathbb{R}: {\rm there\; exists\;} \alpha \in \mathcal{F}(x) \nonumber\\ &{\rm \; such \; that\;}\forall \beta  \in \partial w(x), q = \beta' \alpha\}
\label{LieD}
\end{align}
The above definition of generalized Lie derivative implies that for each $\alpha$ in $\mathcal{F}(x)$, we check if the inner product $\beta'\alpha$ is a fixed value for all $\beta\in\partial w(x)$. If so, this inner product is an element in $\tilde{\mathcal{L}}_{\mathcal{F}}w(x)$, but note that the set $\tilde{\mathcal{L}}_{\mathcal{F}}w(x)$ may be empty.  Moreover, for locally Lipschitz and regular(see \cite{CFHOPNA},p.$39$ and \cite{BACC1999SSD}, p.3 for detailed discussion of regular function\footnote{That a function $w(x):\mathbb{R}^n\to \mathbb{R}$ is called regular at $x\in\mathbb{R}^n$ if\begin{enumerate}
		\item for all $v\in\mathbb{R}^n$ there exists the usual right directional derivative $w_{+}'
		(x,v)$.
		\item for all $v\in\mathbb{R}^n$, $w_{+}'
		(x,v) = w^o(x,v)$.
	\end{enumerate}}) functions $w(x)$, one has the following lemma:
\begin{lemma} (Proposition $10$ in \cite{Cortes08CSM}) \label{Lemma_Gene}
	Let $x:[{0,t_1}]\to\mathbb{R}^r$ be a solution for $\dot x  \in\mathcal{F}(x(t))$, where $\mathcal{F}$ is any set-valued map. Let $w(x): \mathbb{R}^r \to  \mathbb{R}$ be locally Lipschitz and regular. Then $w(x(t))$ is differentiable at almost all $t\in[0,t_1]$; The derivative of $w(x(t))$ satisfies $\frac{{dw(x(t))}}{{dt}} \in\tilde{\mathcal{L}}_{\mathcal{F}}w(x(t)) $ for almost all $t\in[0,t_1]$.
\end{lemma}
\smallskip

Lemma \ref{Lemma_Gene} guarantees the existence of generalized Lie derivatives for functions that are locally Lipschitz and regular. If one focuses on a specific solution, one can show that $\alpha$ in (\ref{LieD}) is a special vector as summarized in the following lemma.
\begin{lemma} (See Proof of Lemma 1 in \cite{BACC1999SSD}) \label{Lemma_LieD}
Let $x(t)$ denote a specific solution of a differential enclosure. Suppose $w(x)$ is locally Lipschitz and regular. Let $\mathcal{I}\subset[0,\infty)$ denote the time interval for which $\dot{x}(t)$ exists. Then \eq{\label{LD2}\frac{dw(x(t)))}{dt}=\beta'\dot x(t)} where $\beta$ is any vector in $\partial w(x)$.
\end{lemma}
%\textcolor{blue}{We used the concept of regular (\cite{CLARK1983NAO}. p.39) in lemma (\ref{Lemma_Gene}). If a function $V:\mathbb{R}^n\to R$ is called regular, then the following two should holds\begin{enumerate}
%	\item for all $v\in \mathbb{R}^n$ there exists the usual right directional derivative $V_{+}'(x,v)$;
%	\item for all $v\in \mathbb{R}^n$, $V'_{+}(x,v) =\lim\limits_{y\to x} \sup\limits_{t\neq 0} \frac{V(y+tv)-V(y)}{t}$.
%\end{enumerate}
%Previous researches \cite{BACC1999SSD} prove in their paper, lemma 1 that if we consider a specific solution x(t) of the differential enclosure, then for a regular and locally Lipschitz function $w(x)$, the generalized Lee-derivative \eq{\label{LD2}\frac{dw(x(t)))}{dt}=\beta'\dot x(t)} where $\beta$ is any vector in $\partial w(x)$. The main argument in their proof is that consider the time when both $w(x)$ and $x(t)$ is differentiable, then from locally Lipschitz they have \eq{\frac{dw(x(t))}{dt}= \lim\limits_{h\to 0}\frac{w(x(t)+h\dot x(t))-w(x(t))}{h}}By letting $h$ tend to $0$ from both side will lead to
%\begin{align}
%	\frac{dw(x(t))}{dt}&&=\max\{\beta'\dot x(t),\beta\in \partial w(x)\} \\ \nonumber&&=\min \{\beta'\dot x(t),\beta\in \partial w(x)\}
%\end{align}}

\subsection{ Preliminary Results} For any positive semi-definite matrix $M\in \mathbb{R}^{r\times r}$, $M\neq 0$ one can define \eq{\label{eq_Phi}\Phi(M)=\{ q\in \mathbb{R}^r|\;\exists \phi  \in F[\sgn (q)],M\phi(q)  = 0\}} and its compliment \eq{\Phi_c(M)=\{ q\in \mathbb{R}^r|\;\forall \phi  \in F[\sgn (q)],M\phi(q)  \neq 0\}.} We impose a further requirement on $M$, namely that $\Phi_c(M)$ is nonempty which can be easily ensured. Let \eq{\Lambda(M)=\{\phi|\phi\in F[\sgn](q),q\in \Phi_c(M)\}} Now $F[\sgn](q)$ is a closed set for any fixed $q$; also note that $F[\sgn](q)$ can only be one of a finite number of different sets; hence it is easy to check for a given $M$ whether $\Phi_c(M)$ is nonempty, (and in a later use of the result, it proves easy to check). It further follows that $\Lambda(M)$ is also a closed set. Consequently, the continuous function $f(\phi)=\phi' M\phi$ has a nonzero minimum on $\Lambda(M)$. We denote \eq{\label{eq_lam}\lambda(M)=\min\limits_{\phi\in\Lambda(M)}f(\phi).}  From the definition of $\Phi_c(M)$ and $\Lambda(M)$, one has $\lambda(M)>0$. To summarize, one has the following lemma:
\begin{lemma} \label{Lemma_lam}
  For any nonnegative-definite matrix $M$, we let $\Phi_c(M)$, $\Lambda(M)$ and $\lambda(M)$ defined as above. Suppose that $\Phi_c(M)$ is nonempty. Then $\lambda(M)$ is a positive constant.
\end{lemma}
\smallskip

For the $m$-node-$\bar{m}$-edge graph $\mathbb{G}$, we label all its nodes as $1,2,...,m$ and all its edges as $1,2,...,\bar{m}$. Assign an arbitrary direction to each edge in $\mathbb{G}$. Then the incidence matrix of $\mathbb{G}$ denoted by $H=[h_{ik}]_{m\times \bar{m}}$ is defined as follows
\eq{h_{ik}=\left\{
             \begin{array}{ll}
               1, & \hbox{$i$ is the head of the $k$th edge;} \\
               -1, & \hbox{$i$ is the tail of the $k$th edge;} \\
               0, & \hbox{otherwise.}
             \end{array}
           \right.
} Since $\mathbb{G}$ is connected, then $\ker H'$ is the span of ${\bf 1 }_m$ \cite{Chung97}. Moreover, one has the following lemma:
\begin{lemma} \label{Lemma_inter}
  Suppose $A$ has full-row rank and $\mathbb{G}$ is connected. Let $\bar{P}=\diag\{P_1,P_2,...,P_m\}$ where each $P_i$ is the projection matrix to $\ker A_i$.  Let $\bar{H}=H\otimes I_n$ with $H$ the incidence matrix of $\mathbb{G}$.  Then one has \eq{\label{KERIM}  \image \bar H\cap \ker \bar P=0} and \eq{\label{eq_p0} \image \bar{H}' \cap \Phi(\bar{H}'\bar{P}\bar{H})=0}
\end{lemma}
\noindent{\bf Proof of Lemma \ref{Lemma_inter}:} Let $u$ be a vector such that \eq{\label{eq_PH}\bar{P}\bar{H} u=0.} The vector $v=\bar H u$ lies in image $\bar H$ and $\ker \bar P$, and we will show it is zero to establish (\ref{KERIM}). Define $\bar A=\diag \{A_1,A_2,...,A_m\}$, which is full row rank since $A$ has full row rank. Then $\bar P=I-\bar A'(\bar A \bar A')^{-1}\bar A$. It follows from (\ref{eq_PH}) that \begin{equation} \label{ki2}
		\bar H u=\bar A'(\bar A \bar A')^{-1}\bar A \bar H u
		\end{equation} Multiplying both sides of the above equation by ${\textbf 1}_m'\otimes I_n$, one has
		\begin{equation} \label{ki3}
		0=({\textbf 1_m\otimes I_n})' \bar A'(\bar A \bar A')^{-1}\bar A \bar H u
		\end{equation} Since $({\textbf 1}_m'\otimes I_n' )\bar A'=A'$ there holds
		\begin{equation} \label{ki4}
		0= A'(\bar A \bar A')^{-1}\bar A \bar H u
		\end{equation}
		Since $A'$ is full column rank, one has
		\begin{equation} \label{ki5}
		0= (\bar A \bar A')^{-1}\bar A \bar H u
		\end{equation}
		From (\ref{ki2}) and equation (\ref{ki5}) one has
		\begin{equation} \label{ki6}
		\bar H u=0
		\end{equation} Furthermore, we notice (\ref{eq_PH}) and conclude that (\ref{KERIM}) is true.
\smallskip

%\begin{lemma}\label{Lemma_p0}
%Suppose $A$ is with full-row rank and $\mathbb{G}$ is connected. Then
%\end{lemma}
%\noindent{\bf Proof of Lemma \ref{Lemma_p0}:}

For any $q\in \image \bar{H}' \cap \Phi(\bar{H}'\bar{P}\bar{H})$, there exists a vector $p$ such that \eq{\label{eq_qhp}q=\bar{H}'p} and a vector $\phi  \in F[\sgn] (q)$ such that $\bar{H}'\bar{P}\bar{H}\phi=0$. Note that $\bar{P}$ is a projection matrix; then \eq{\bar{P}\bar{H}\phi=0. \label{eq_HPH}} From (\ref{KERIM}) one then has \eq{\bar{H}\phi=0.\label{eq_bhp}} From $\phi  \in F[\sgn] (q)$ and (\ref{eq_xl1}), one has $$||q||_1=\phi'q$$ which together with (\ref{eq_qhp}) and (\ref{eq_bhp}) implies $||q||_1=0$. Then one has $q=0$ and (\ref{eq_p0}) is true. $\qed$
\smallskip

Now consider the system \eq{\dot{x}\in -MF[\sgn](x)\label{eq_sys0}} with any positive semi-definite matrix $M\in \mathbb{R}^{r\times r}$. The existence of a Filippov solution to (\ref{eq_sys0})  can be guaranteed by Lemma \ref{Lemma_Fili}. The existence interval is $t\in [0,\infty)$ because of the global bound on the right-hand side to (\ref{eq_sys0}). Let $x(t)$ denote such a Filippov solution for any given $x(0)$.  Note that the function ${\|x\|}_{1}$ is locally Lipschitz and regular(The word was introduced early with reference). Then by Lemma \ref{Lemma_Gene}, the time derivative of $\|x(t)\|_1$ exists for almost all $t\in [0,\infty)$  and is in the set of generalized Lie derivatives. In other words, there exists a set $\mathcal{I}=[0,\infty)\setminus \mathcal{T}$ with $\mathcal{T}$ of Lebesgue measure 0 such that
\begin{equation}\label{LDrive0}
\frac{{d\|x(t)\|_1}}{{dt}} \text{ exists for all } t\in \mathcal{I}
\end{equation}
\begin{proposition}\label{Propo}
 Let $x(t)$ denote a Filippov solution to (\ref{eq_sys0}) for any given $x(0)\in \mathbb{R}^r$. Then
\begin{itemize}
  \item \begin{equation}\label{eq_nonincrease}
\frac{{d\|x(t)\|_1}}{{dt}} \leq 0, \quad t\in  \mathcal{I};
\end{equation}
  \item  there exists a finite time $T$ such that \eq{x(T)\in \Phi(M);}
  \item  if it is further true that \begin{equation}\label{Assum}
\frac{{d\|x(t)\|_1}}{{dt}} =0, \quad t\in [T,\infty)\setminus \mathcal{T},
\end{equation}  one has \eq{x(t)=x(T)\quad t\in [T,\infty) \label{eq_propo3}}
\end{itemize}

\end{proposition}
\smallskip
\begin{remark}
Note that the right-hand side of (\ref{eq_sys0}) is a projection of the gradient flow of the potential function $\|x(t)\|_1$. It is also standard that the gradient law of a real analytic function with a lower bound converges to a single point which is both a local minimum and a critical point of the potential function \cite{APKKSCL}. However, if the real analytic property does not hold, the convergence result may fail. Indeed, the function $\|x(t)\|_1$ here is obviously not real analytic, and one cannot immediately assert that (\ref{eq_sys0}) will drive $\|x(t)\|_1$ to its minimum, not to mention the finite time result in Proposition \ref{Propo}. Thus Proposition \ref{Propo} is nontrivial and will serve as the foundation for devising finite-time distributed linear equation solvers in this paper.
\end{remark}
\smallskip

\noindent{\bf Proof of Proposition \ref{Propo}:} By (\ref{LD2}) in Lemma \ref{Lemma_LieD}, one has
\begin{equation}\label{LDrive}
\frac{{d||x(t)||_1}}{{dt}} =\beta'\dot{x},\quad t\in \mathcal{I}
\end{equation} holds for all $\beta\in\partial{\|x(t)\|_1}$. It follows from $\dot{x}\in -MF[\sgn](x)$ in (\ref{eq_sys0}) that at each $t$ there exists a $\gamma(t)\in{F[\sgn](x)}$ such that \eq{\label{eq_gamma}\dot{x}=-M\gamma(t)} Since $\beta$ could be chosen as any vector in $\partial \|x(t)\|_1$, $\gamma(t)\in F[\sgn](x)$ and $\partial \|x\|_1=F[\sgn](x)$ from (\ref{equa}),  one could choose $\beta=\gamma(t)$ in (\ref{LDrive}), which together with (\ref{eq_gamma}) leads to
\begin{equation}
\frac{{d\|x(t)\|_1}}{{dt}}=-\gamma(t)'M\gamma(t),\quad t\in \mathcal{I}
\label{Con1}
\end{equation} Note that $M$ is positive semi-definite. Thus (\ref{eq_nonincrease}) is true.

We use the method of contradiction to prove that there exists a finite time $T$ such that $x(T)\in \Phi(M),\ T\in \mathcal{I}.$ Suppose such a finite time does not exist. One has $$x(t)\in \Phi_c(M),\quad t\in \mathcal{I}$$ Then $$\frac{{d\|x(t)\|_1}}{{dt}}\leq -\lambda(M),\quad t\in \mathcal{I}$$ where $\lambda(M)$ is as defined in (\ref{eq_lam}). It follows that $$\|x(t)\|_1\leq \|x(0)\|_1-\lambda(M) t,\quad t\in [0,\infty).$$ This contradicts the fact that $\|x(t)\|_1\geq 0, t\in [0,\infty)$ since $\lambda(M)$ is a  positive constant by Lemma \ref{Lemma_lam}. Thus there exists a finite time $T$ such that $x(T)\in \Phi(M)$.

From the assumption (\ref{Assum}), the fact that $M$ is semipositive definite and (\ref{Con1}), one has \eq{M\gamma(t)=0,\quad t\in [T,\infty)\setminus \mathcal{T}.} It follows from this and (\ref{eq_gamma}) that $$\dot{x}(t)=0,\quad t\in [T,\infty)\setminus\mathcal{T}.$$ By integration of $\dot x(t)$ from $T$ to $\infty$, one has $x(t)=x(T),\quad t\in [T,\infty)$. This completes the proof. $\qed$

\section{Algorithms and Main Results}\label{sect4}
In this section, we study three related problems: finite time distributed solution of $Ax=b$, centralized solution of $Ax=b$ achieving a minimum $l_1$-norm, and then finally, using ideas from the first two, and finding a distributed algorithm achieving the minimum $l_1$-norm solution of $Ax=b$ in finite time.
\subsection{Finite-Time Distributed Linear Equation Solver}
In this subsection, we will present a distributed update for achieving a solution $x^*$ to $Ax=b$ in finite time. Of course, $A$ is assumed to have full row rank, but not necessarily be square. Recall that distributed linear equation solvers based on the agreement principle require each agent $i$ to limit its update subject to its own constraint $A_ix=b_i$ while seeking consensus with its neighbors \cite{SJA15TAC}. Such an agreement principle in continuous-time systems can be achieved by a projection-consensus flow which at each agent $i$ projects a function of its neighbors' states to the subspace defined by its linear equation \cite{GB16ACC}. By choosing the finite-time gradient flow for consensus developed by \cite{J2006Auto} within the projection-consensus flow, we are led to postulate the following update for each agent $i$:
 \eq{\label{update1}\dot y_i= -P_i\sum_{j\in\mathcal{N}_i}\phi_{ij},\quad A_iy_i(0)=b_i} where $\phi_{ij}\in F[\sgn](y_i-y_j)$ and $P_i$ denote the projection matrix to the kernel of $A_i$.

Note the special property that $F[\sgn](0)$ is a point which can be chosen arbitrarily from the interval $[-1,1]$. Generally speaking, different agents may have different choices for $F[\sgn](0)$. Before proceeding, we make the following assumption on coordinations between neighbor agents:
\begin{assumption}\label{Assum1}
Each pair of neighbor agents $i$ and $j$ takes the choice of $\phi_{ij}$ and $\phi_{ji}$ when $y_i=y_j$ such that \eq{\label{basic}\phi_{ij}=-\phi_{ji}}
\end{assumption}
Under Assumption \ref{Assum1} and the definition of $F[\sgn](x)$, one always has $\phi_{ij}=-\phi_{ji}$ no matter whether $y_i$ is equal to $y_j$ or not.

%\textcolor{blue}{\textbf{remark}: $F[\sgn](0)$ is a special point in our algorithm. Cause the derivative can take any value from the interval $[-1,1]$. And for a vector $x\in\mathbb{R}^n$, if there are two entry of $x$ are zero, $F[\sgn](x)$ can take different value at those two entries. However, we require the value of $F[\sgn](y_i-y_j)$ taken in agent $i$ to be the opposite of $F[\sgn](y_j-y_i)$ in agent $j$, which is \eq{\label{basic}F[\sgn](y_i-y_j)+F[\sgn](y_j-y_i)=0} we call this \textbf{basic control law}. Since our graph is undirected, then the \textbf{basic control law} does not affect the distributed nature of out algorithm.}

Let $y=\col\{y_1,y_2,\cdots,y_m\}$, $\bar P={\rm diag}\{P_1,P_2,\cdots,P_m\}$ and $\bar H=H\otimes I_n$ with $H$ the incidence matrix of $\mathbb{G}$.  Then from (\ref{update1}) we have \eq{\label{update1b}\dot y\in -\bar{P} \bar H F[\sgn](\bar H'y)} By Lemma \ref{Lemma_Fili} there exists a Filippov solution to system (\ref{update1b}) for $t\in [0,\infty)$, we denote this solution by $y(t)$. By Lemma \ref{Lemma_Gene}, there exists a set $\mathcal{I}=[0,\infty)\setminus \mathcal{T}$ with $\mathcal{T}$ of Lebesgue measure 0 such that $\frac{{d\|y(t)\|_1}}{{dt}} $ exists for all $ t\in \mathcal{I}$.
Moreover, one has the following main theorem, which establishes existence of a limiting consensus solution but for the moment leaves unspecified the $L_1$-optimality.
\begin{theorem} \label{T1}
Under Assumption \ref{Assum1} and the updates (\ref{update1}), and with $A$ assumed to have full row rank, all $y_i(t)$, $i=1,2,...,m$, converge to be a single solution to  $Ax=b$ in finite time.
\end{theorem}
\bigskip

\noindent{\bf Proof of  Theorem \ref{T1}.} Since $A_iy_i(0)=b_i$ and $\dot{y}_i$ is in the kernel of $A_i$, one has $A_iy_i(t)=b_i$ for all $t\in [0,\infty)$.  Then to prove Theorem \ref{T1}, it is sufficient to show that all $y_i$ reach consensus in finite time. Note that $\mathbb{G}$ is connected and $\ker H'$ is spanned by the vector ${\bf 1}_m$.  Then $\bar{H}'y=0$ if and only if all $y_i$ are equal. Thus to prove Theorem \ref{T1}, it suffices to prove $z(t)=\bar{H}'y(t)$ converges to 0 in finite time. By multiplying both sides of (\ref{update1b}) by $\bar H'$, one has \eq{\label{update1c}\dot{z}\in-\bar H'\bar P\bar H F[\sgn](z)}

By Proposition \ref{Propo}, we have \begin{equation}\label{eq_nonincrease1}
\frac{{d\|z(t)\|_1}}{{dt}} \leq 0, t\in  \mathcal{I};
\end{equation} and there exists a finite time $T\in \mathcal{I}$ such that \eq{z(T)\in \Phi(\bar H'\bar P\bar H).} From this, we know the fact that $z(T)=\bar {H}'y(T)$ so that  $z(T)\in \image \bar{H}'$, and recalling (\ref{eq_p0}) in Lemma \ref{Lemma_inter}, one has  $$z(T)=0,$$ which by (\ref{eq_nonincrease1}) implies  $$\|z(t)\|_1=0,\quad t\in [T,\infty).$$ It follows that $$z(t)=0, \quad t\in [T,\infty).$$ This completes the proof. $\qed$

\subsection{Finite-Time Centralized Update for Minimum $l_1$-norm Solution}\label{sec2}
In this subsection, we will propose a centralized update for achieving the minimum $l_1$-norm solution to $Ax=b$. By noting that $\|x\|_1$ is convex, we conceive of using a negative gradient flow of $||x||_1$ subject to $x$ remaining on the manifold $Ax=b$ in order to achieve $\bar{x}^*=\arg\min_{Ax=b}  \|x\|_{1}$ . This leads us to the following update:
\begin{equation}
\dot{y} \in-PF[\sgn](y),\quad Ay(0)=b.
\label{l1a}
\end{equation} where $P$ denotes the projection matrix onto the kernel of $A$. Again by Lemma \ref{Lemma_Fili} one has there exists a Filippov solution to system (\ref{l1a}) for $t\in [0,\infty)$, which we denote by $y(t)$. By Lemma \ref{Lemma_Gene}, there exists a set $\mathcal{I}=[0,\infty)\setminus \mathcal{T}$ with $\mathcal{T}$ of measure 0 such that $\frac{{d\|y(t)\|_1}}{{dt}} $ exists for all $ t\in \mathcal{I}$. Moreover, we have the following main theorem:
\begin{theorem} \label{T2a}
With $A$ of full row rank, the Filippov solution $y(t)$ to (\ref{l1a}) converges in finite time to a constant, which is the minimum $l_1$-norm solution to $Ax=b$.
\end{theorem}
\bigskip

\noindent{\bf Proof of Theorem \ref{T2a}:} By Proposition \ref{Propo}, one has \begin{equation}\label{eq_nonincrease2}
\frac{{d\|y(t)\|_1}}{{dt}} \leq 0, t\in  \mathcal{I};
\end{equation} and there exists a finite time $T\in \mathcal{I}$ such that $$y(T)\in \Phi(P).$$ Then there exists a vector $\phi \in F[\sgn](y(T))$ such that $P\phi=0$. This and (\ref{eq_xl1}) imply
\begin{equation}\label{abseq1}
	\phi'y(T)=\|y(T)\|_1
\end{equation} Moreover, let $\bar{y}$ denote any solution to $Ax=b$. Recall (\ref{ineq}), there holds \begin{equation}\label{abseq11}
	\phi'\bar{y}\leq \|\bar{y}\|_1
\end{equation}
Since $\phi \in \ker P$, one has $\phi \in \image A'$. This implies that there exists a vector $q$ such that
\begin{equation}\label{eq_zA}
	q'A=\phi'
\end{equation} From $Ay(T)=b=A\bar{y}$ and (\ref{abseq1})-(\ref{eq_zA}), one has
\begin{equation}
\|y(T)\|_1=\phi'y(T)=q'Ay(T)=q'A\bar{y}=\phi'\bar{y}\leq \|\bar{y}\|_1
\end{equation}
where $\bar{y}$ is any solution to $Ax=b$. Thus $y(T)$ is a minimum $l_1$ norm solution to $Ax=b$. This and (\ref{eq_nonincrease2}) implies $||y(t)||_1$ reaches its minimum value subject to $Ay(t)=b$ for $t\in [T,\infty)\setminus \mathcal{T}$. Thus \begin{equation}
\frac{{d\|y(t)\|_1}}{{dt}}=0,\quad t\in [T,\infty)\setminus \mathcal{T}
\label{Con3}
\end{equation} which satisfies the assumption (\ref{Assum}) in Proposition \ref{Propo} again. Then $y(t)=y(T),\ t\in [T,\infty)$. Thus $y(t)$  is the minimum $l_1$-norm solution to $Ax=b$ for $t\in [T,\infty)$. This completes the proof. $\qed$

\subsection{Finite-Time Distributed Update for Minimum $l_1$-norm Solutions}\label{sec3}
In this subsection we will develop a distributed update for a multi-agent network to achieve the minimum $l_1$-norm solution to $Ax=b$ in finite time. Motivated to study a combination of the finite-time distributed linear equation solver in (\ref{update1}) and the finite-time centralized update for minimum $l_1$-norm solutions in (\ref{l1a}), we propose the following update for agent $i$, $i=1,2,...,m$:
\begin{align}
\dot{y}_i = - k(t){P_i}\phi_i - {P_i}\sum\limits_{j \in {\mathcal{N}_i}} \phi_{ij}
\label{l1b}
\end{align} where $\phi_i\in F[\sgn] (y_i)$, $\phi_{ij}\in F[\sgn]({y_i} - {y_j})$, with $\phi_{ij}=-\phi_{ji}$ in case $y_i=y_j$, and
\eq{ \label{initial}A_iy_i(0)=b_i.} We assume that $k(t)\in \mathbb{R}$ is measurable and locally bounded almost everywhere for $t\in [0,\infty)$, and
\begin{equation} \label{kt1}
\lim\limits_{t\to\infty}k(t)=\delta
\end{equation}
\begin{equation} \label{kt2}
\int_{0}^{\infty}k(t)dt=\infty
\end{equation}
where $\delta$ is a sufficiently small nonnegative number depending on the connection of the network and $A$, note that $0$ is always a feasible choice of $\delta$. One example of a choice of $k(t)$ is $k(t)=\frac{\bar \delta}{t+1}+\delta$. One simple case is choosing $\bar \delta=1$, $\delta=0$, and resulting in $k(t)=\frac{1}{t+1}$, obtained by taking $\delta$ to be zero. This choice obviates the need to decide how small one has to be to meet a ``sufficiently small" condition, but may result in rather slow convergence. Now from $A_iy_i(0)=b_i$ and the fact that $P_i$ is the projection to the kernel of $A_i$ (which ensures $\dot y_i\in \ker A_i$), one has
%\and there exists $t_1,t_2,...$ such that $[0,\infty )=[0,t_1)\cup [t_1,t_2)\cup \cdots $ and $k(t)$ is a positive constant in each of the interval.

\begin{equation}\label{stateeq1}
 	A_iy_i(t)=b_i, \quad t\in [0,\infty)
 \end{equation}

Let $y=\col \{y_1,y_2,y_3,...,y_m\}$, $\bar{P}=\diag\{P_1,P_2,...,P_m\}$, and $\bar H=H\otimes I_n$ with $H$ the incidence matrix of $\mathbb{G}$. From the updates in (\ref{l1b}) and Assumption \ref{Assum1}, one has
		\begin{equation} \label{l1c}
		\dot y\in- k(t) \bar P  F[\sgn](y) -\bar P \bar H F[\sgn](\bar H'y)
		\end{equation} Note that  $\sgn(y)$, $k(t)$ and $\bar P \bar H F[\sgn](\bar H'y)$ are measurable and locally bounded almost everywhere. Then by Lemma \ref{genfilip} there exists a Filippov solution to system (\ref{l1c}) for any given $y(0)$ satisfying (\ref{initial}), which we denote by  $y(t)=\col \{y_1(t),y_2(t),y_3(t),...,y_m(t)\}$ .  By Lemma \ref{Lemma_Gene}, there exists a set $\mathcal{I}=[0,\infty)\setminus \mathcal{T}$ with $\mathcal{T}$ of Lebesgue measure 0 such that $\frac{{d\|y(t)\|_1}}{{dt}} $ exists for all $ t\in \mathcal{I}$. Then one has the following theorem:

\begin{theorem} \label{T3}
Under Assumption \ref{Assum1} and the update (\ref{l1b}) and with $A$ of full row rank, all $y_i(t)$, $ i=1,2,...,m$ converge in finite time to the same value which is the minimum $l_1$-norm solution to $Ax=b$.
\end{theorem}
\bigskip

\noindent{\bf Proof of Theorem \ref{T3}:} We first prove that all $y_i(t)$  reach a consensus in finite time by showing that $z(t)$ converges to 0 in finite time, where $z(t)=\bar H'y(t)$. Multiplying both sides of (\ref{l1c}) by $\bar{H}'$, one has
		\begin{equation} \label{up1}
		\dot{z} \in- k(t) \bar H' \bar P  F[\sgn](y) -\bar H' \bar P \bar H F[\sgn](z)
		\end{equation} By Lemma \ref{Lemma_LieD}, one has $$\frac{d\|z(t)\|_1}{dt}=\beta'\dot{z},\quad t\in \mathcal{I}$$ where $\beta$ can be any vector in $\partial \|z\|_1$. Note that $\partial \|z\|_1=F[\sgn](z)$. Then \eq{\label{dv}\frac{d\|z(t)\|_1}{dt}= -k(t)\gamma' \bar H' \bar P\eta-\gamma'\bar H' \bar P \bar H\gamma,\quad t\in \mathcal{I}} where $\eta\in F[\sgn](y)$, $\gamma \in F[\sgn](z)$ and $\beta$ is chosen to be equal to $\gamma$.
Since $z(t)\in \image \bar{H}'$, then by Lemma \ref{Lemma_inter}, if also $z(t)\in \Phi(\bar{H}'\bar{P}\bar{H})$, one will have $z(t)=0$. Thus \eq{z(t)\in \Phi_c(\bar{H}'\bar{P}\bar{H}) \text{ as long as } \|z(t)\|_1\neq 0.} Now by the definition of $\lambda(\bar{H}'\bar{P}\bar{H})$ and Lemma \ref{Lemma_lam}, one has \eq{\label{eq_bound}\gamma'\bar H' \bar P \bar H\gamma\geq \rho \text{ as long as } \|z(t)\|_1\neq 0} where $\rho=\lambda(\bar H' \bar P \bar H)$ is a positive constant. Let $\kappa(\bar H, \bar P)$ denote an upper bound on $|\gamma'\bar H'\bar P\eta|$, and define an upper bound on $\delta$ by \eq{\label{delta} \delta <\frac{\rho}{\kappa(\bar H,\bar P)}} This captures the idea stated previously that $\delta$ depends on $A$ and the graph. For any $\delta$ chosen as in (\ref{delta}), there must exist a finite time $T_1$ such that \eq{\label{eq_finiteT}|-k(t)\gamma \bar H' \bar P\eta|\leq \rho_2, \quad t\in [T_1,\infty).} where we have $\rho_2=\kappa(\bar H,\bar P)\delta<\rho$. From (\ref{dv}), (\ref{eq_bound}) and (\ref{eq_finiteT}), one has \eq{\label{minrho}\frac{d\|z(t)\|_1}{dt}\leq -(\rho-\rho_2)\text{ as long as } \|z(t)\|_1\neq 0, \ t\in [T_1,\infty)\setminus \mathcal{T}  } with $\rho$ a positive constant. Thus there must exist a finite time $T_2\geq T_1$ such that
\eq{\label{eq_zfinite}z(T_2)=0}

Next we prove that \eq{\label{eq_zfinite2}z(t)=0,\quad t\in [T_2,\infty)} We prove this by contradiction. Suppose (\ref{eq_zfinite2}) is not true. Then there exists a time $\bar{T}_2>T_2$ such that $z(\bar{T}_2)\neq 0$. Then $\|z(\bar{T}_2)\|_1>0$. Since $\|z(t)\|_1$ is continuous, there exists a time $T_2^*$ such that $\|z(T_2^*)\|_1$ takes its maximum value for $t\in [T_2,\bar{T}_2]$. Again, since $\|z(t)\|_1$ is continuous, there exists a sufficiently small but positive $\epsilon$ such that $\|z(t)\|_1>0$ is differentiable for $t\in [T_2^*-\epsilon, T_2^*]$. Because $\|z(t)\|_1$ is differentiable almost everywhere, we know that\eq{\label{int1}\|z(T_2^*)\|_1=\int_{T_2^*-\epsilon}^{T_2^*}\frac{d\|z(t)\|_1}{dt}dt+\|z(T_2^*-\epsilon)\|_1} Because $\|z(t)\|_1>0$ in $[T_2^*-\epsilon,T_2^*]$, by (\ref{minrho}) and (\ref{int1}) we have \eq{\label{int2}\|z(T_2^*)\|_1\leq-\epsilon(\rho-\rho_2)+\|z(T_2^*-\epsilon)\|_1<\|z(T_2^*-\epsilon)\|_1} This contradicts the fact that $\|z(T_2^*)\|_1$ is the maximum value on $[T_2,\bar{T}_2]$. Thus (\ref{eq_zfinite2}) is true.

By (\ref{eq_zfinite2}), one has there exists a vector $\bar{y}(t)$ such that \eq{\label{eq_ycon}y_1(t)=y_2(t)=\cdots =y_m(t)=\bar{y}(t),\quad t\in [T_2,\infty)} Moreover, $A\bar{y}(T_2)=b$ since $A_iy_i(t)=b_i$ for $i=1,2,...,m$. To prove Theorem \ref{T3}, we only need to prove that $\bar{y}(t)$ converges to be the minimum $l_1$-norm solution to $Ax=b$. To see why this is so, we let $P$ denote the projection matrix to the kernel of $A$. Then $ P P_i= P$ for $i=1,2,...,m$.  Multiplying both sides of (\ref{l1b}) by $P$, one has \eq{\label{l1d0}\dot {y}_i= -k(t)P \phi_i-P\sum_{j\in\mathcal{N}_i}\phi_{ij},\ i=1,2,...,m} Since $\mathbb{G}$ is undirected, then $\phi_{ij}$ appears in the update $i$ if $\phi_{ji}$ appears in its neighbor $j$'s update. By adding the updates in (\ref{l1d0}) for $i=1,2,...,m$ and noting $\phi_{ij}=-\phi_{ji}$ for any two neighbors $i$ and $j$, one has \eq{\label{l1d1}\sum_{i=1}^m\dot {y}_i= -k(t)P \sum_{i=1}^m\phi_i} where $\phi_i\in F[\sgn](y_i)$. By (\ref{eq_ycon}), one knows all $y_i(t)$ reach a consensus $\bar{y}(t)$ for $t\in [T_2,\infty)$. Note that if the $k$th entry of $\bar{y}(t)$ is 0, the $k$th entry of each $\phi_i$ can be selected as an arbitrary value from $[-1,1]$, which may be different for different entries, but their average is still an arbitrary value in $[-1,1]$. Thus \eq{\label{l1s}\frac{1}{m}\sum_{i=1}^m \phi_i\in F[\sgn](\bar{y}(t)),\quad t\in [T_2,\infty)} From (\ref{l1d1}) and (\ref{l1s}) we have
\eq{\label{l1e2}\dot {\bar{y}}\in -k(t) P F[\sgn](\bar{y}),\quad t\in [T_2,\infty)}
Let $\tau=\int_{T_2}^{t}k(s)ds$. From $$\frac{d\bar{y}}{d\tau}=\frac{d\bar{y}}{dt}\cdot \frac{dt}{d\tau}$$ and (\ref{l1e2}), one has
	\eq{\label{l1g}\frac{d\bar{y}}{d\tau}\in -PF[\sgn](\bar{y}),\quad \tau\in [0, \infty)} with $A\bar{y}(\tau)=b$ for $\tau=0$. This is exactly the same as the centralized update in (\ref{l1a}). By Theorem \ref{T1}, there exists a finite time $\Gamma$ such that $y(\tau)$ is the minimum $l_1$-norm solution to $Ax=b$ for $\tau\in [\Gamma, \infty)$. By the relation between $\tau$ and $t$, one has correspondingly that there exist a finite time $T$ such that $\bar{y}(t)$ is a minimum $l_1$-norm solution for $t\in [T,\infty)$. This completes our proof. $\qed$
\section{Simulation Result}\label{sect5}
\begin{figure}[th]
	\centering
	\includegraphics[width=2.2 in]{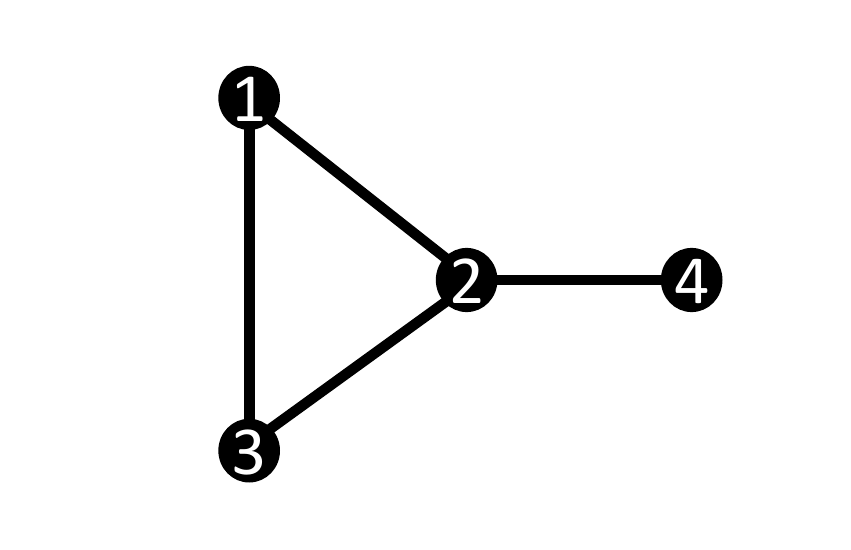} \caption{A four agent network} \label{f1}
\end{figure}

In this section, we will report several simulations of the proposed algorithms for solving an underdetermined linear equation $Ax=b$ in a four-agent undirected and connected network as in Figure \ref{f1}.

\noindent{Here,} $A$ and $b$ are partitioned as $A=\matt{A_1'&A_2'&A_3'&A_4'}'$ and $b=\matt{b_1'&b_2'&b_3'&b_4'}'$, respectively. Each agent $i$ knows $A_i$ and $b_i$ with
\footnotesize
\begin{align}
&A_1'=\matt{0.63&0.04\\
		     0.58&0.60\\
		     0.65&0.50\\
		     0.33&0.81\\
		     0.68&0.01\\
		     0.22&0.51\\
		     0.49&0.23\\
		     0.21&0.29\\
		     0.62&0.25\\
		     0.57&0.21\\
		     0.71&0.66\\
		     0.28&0.90},
    \quad A_2'=\matt{0.80&0.99\\
		     0.25&0.65\\
		     0.53&0.38\\
		     0.79&0.12\\
		     0.13&0.76\\
		     0.79&0.52\\
		     0.34&0.55\\
		     0.45&0.24\\
		     0.10&0.55\\
		     0.94&0.51\\
		     0.78&0.58\\
		     0.70&0.85}
\\ \nonumber &A_3'=\matt{0.44&0.34\\
		     0.06&0.94\\
		     0.77&0.28\\
		     0.16&0.41\\
		     0.84&0.75\\
		     0.62&0.56\\
		     0.74&0.41\\
		     0.26&0.89\\
		     0.44&0.69\\
		     0.28&0.23\\
		     0.50&0.88\\
		     0.38&0.63},
\quad	 A_4'=\matt{0.05&0.23\\
		     0.09&0.33\\
		     0.65&0.92\\
		     0.69&0.66\\
		     0.94&0.92\\
		     0.73&0.06\\
		     0.51&0.13\\
		     0.59&0.94\\
		     0.76&0.40\\
		     0.95&0.69\\
	         0.39&0.24\\
		     0.03&0.92}
\end{align}
%	A'=\matt{0.63&0.04&0.80&0.99&0.44&0.34&0.05&0.23\\
%		0.58&0.60&0.25&0.65&0.06&0.94&0.09&0.33\\
%		0.65&0.50&0.53&0.38&0.77&0.28&0.65&0.92\\
%		0.33&0.81&0.79&0.12&0.16&0.41&0.69&0.66\\
%		0.68&0.01&0.13&0.76&0.84&0.75&0.94&0.92\\
%		0.22&0.51&0.79&0.52&0.62&0.56&0.73&0.06\\
%		0.49&0.23&0.34&0.55&0.74&0.41&0.51&0.13\\
%		0.21&0.29&0.45&0.24&0.26&0.89&0.59&0.94\\
%		0.62&0.25&0.10&0.55&0.44&0.69&0.76&0.40\\
%		0.57&0.21&0.94&0.51&0.28&0.23&0.95&0.69\\
%		0.71&0.66&0.78&0.58&0.50&0.88&0.39&0.24\\
%		0.28&0.90&0.70&0.85&0.38&0.63&0.03&0.92}
%\eq{b'=\matt{0.47&0.52&0.77&0.34&0.63&0.33&0.31&0.65}}
\begin{align}
&b_1'=\matt{0.47&0.52},\quad b_2'=\matt{0.77&0.34}\\ \nonumber&
b_3'=\matt{0.63&0.33},\quad b_4'=\matt{0.31&0.65}
\end{align}
\normalsize

\smallskip

\begin{figure}[th]
		\centering
		\includegraphics[width=3.7 in]{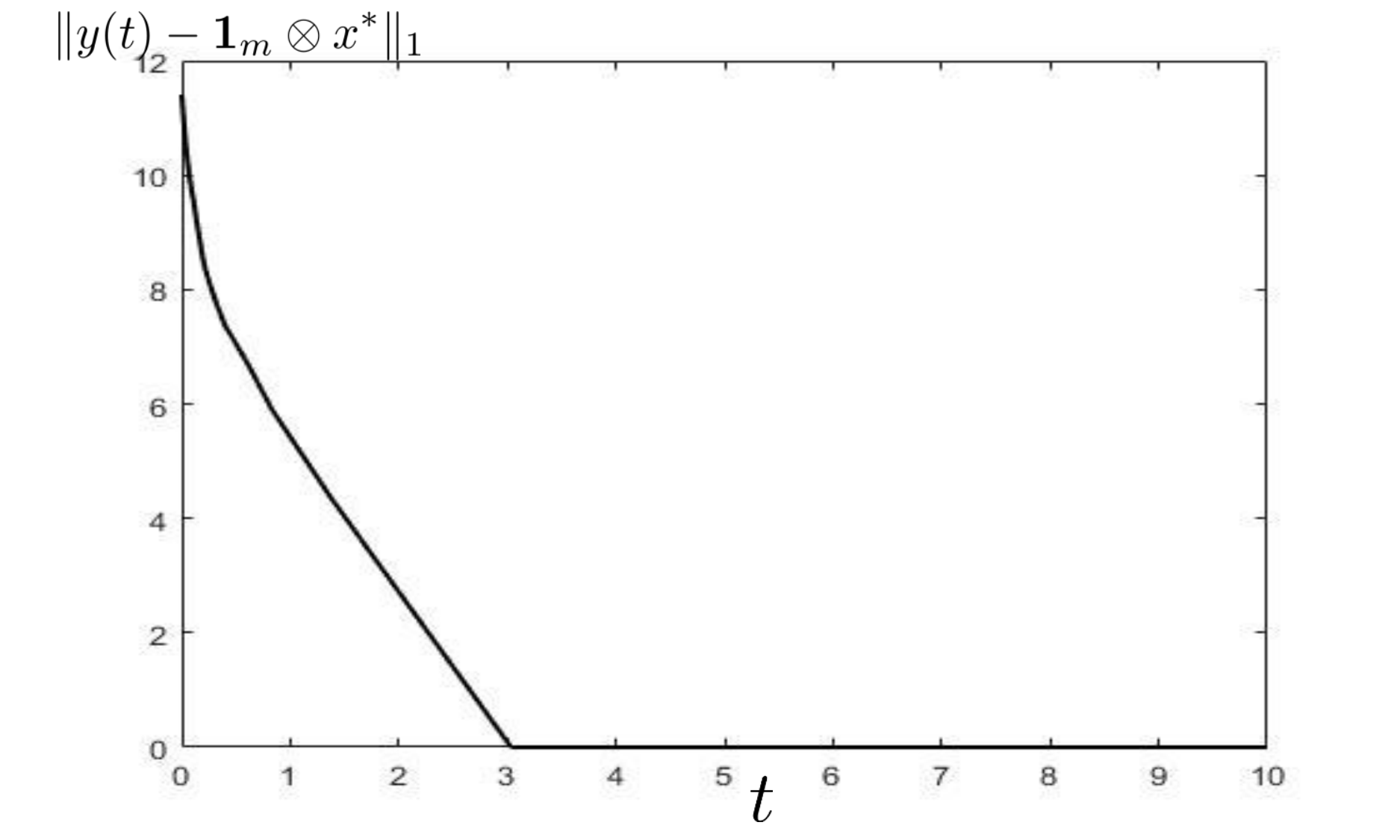}  \caption{Finite time achieving a solution under the update(\ref{update1})}\label{f2}
	\end{figure}

\textbf{Example 1}: We utilize the distributed update (\ref{update1}) to achieve a solution to $Ax=b$ denoted by $x^*$ in finite time in the above four-agent network. Let $y=\matt{y_1'&y_2'&y_3'&y_4'}'$ where $y_i(t)$ denote the state of agent $i$ that is the estimate of agent $i$ to $x^*$. Then $\|y(t)-\textbf{1}_m \otimes x^*\|_1$ measures the difference between all agents' estimations and the solution $x^*$.
As shown by simulations in Figure \ref{f2}, $\|y(t)-\textbf{1}_m \otimes x^*\|_1$ reaches $0$ in finite time. This suggests all agents' states achieves a consensus to $x^*$ in finite time, consistent with the claim of Theorem \ref{T1}.

\begin{figure}[th]
		\centering
		\includegraphics[width=3.7 in]{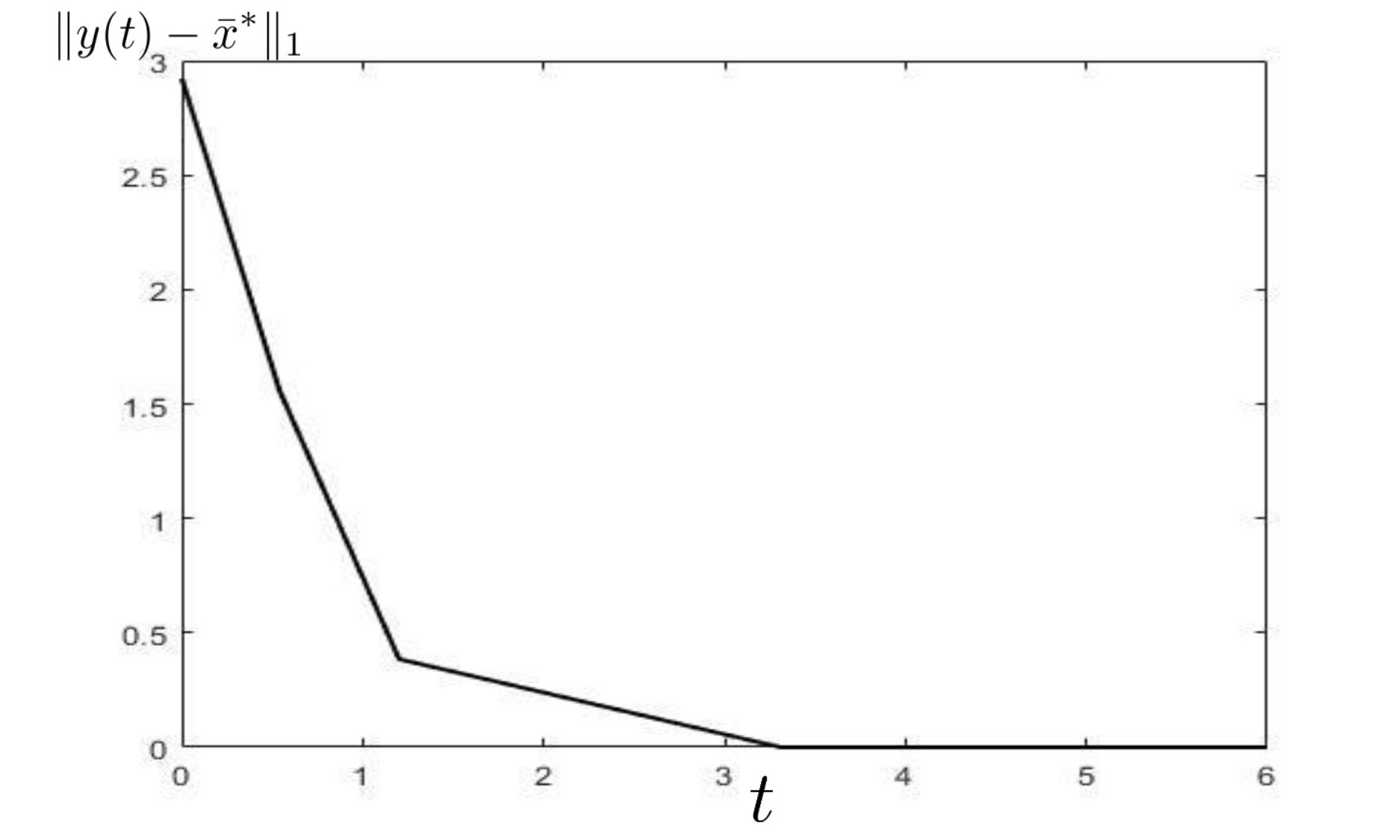}  \caption{Centralized solver for achieving a minimum $l_1$ norm solution under the update (\ref{l1a})}\label{f3}
	\end{figure}

\textbf{Example 2}: We employ the centralized update (\ref{l1a}) with state vector $y(t)$ to achieve $\bar x^*$ which denotes a minimum $l_1$-norm solution to $Ax=b$. As shown in Figure \ref{f3}, $\|y(t)-\bar x^*\|_1$ reaches $0$ in finite time and maintains to be $0$ afterwards. This indicates that the minimum $l_1$-norm solution $\bar{x}^*$ is achieved in finite time corresponding to Theorem \ref{T2a}. It is worth noting that one could observe multiple phases of convergence in Figure \ref{f2}. This is because $F[\sgn](y(t))$ in the update (\ref{l1a}) takes different values piece-wisely, and results in different convergence rates.

\begin{figure}[th]
		\centering
		\includegraphics[width=3.7 in]{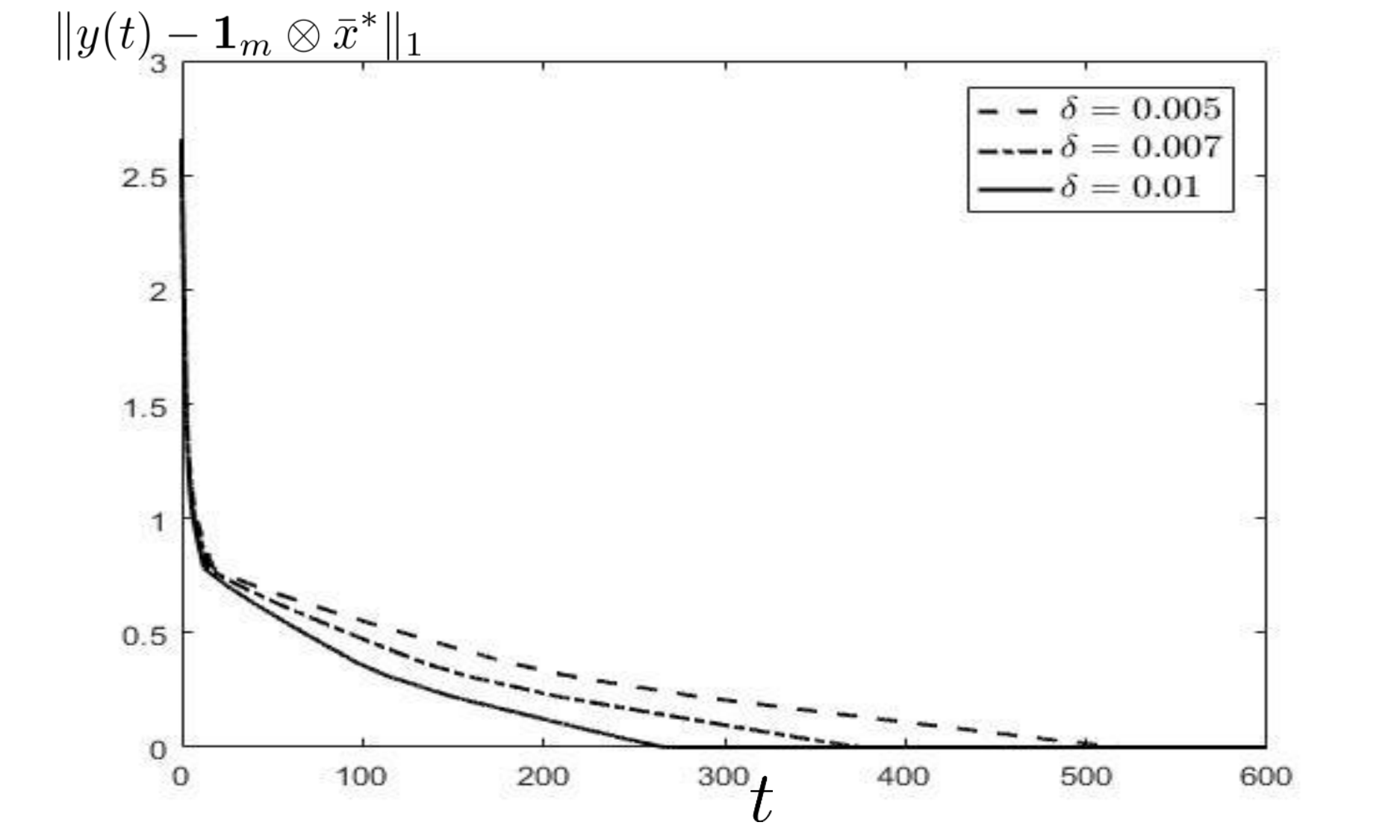}  \caption{Distributed solver for achieving minimum $l_1$ norm solution under the update (\ref{l1b}), where $k(t)=\frac{\bar{\delta}}{t+1}+\delta$ with fixed $\bar \delta=0.1$ and different values of $\delta$.}\label{f4}
	\end{figure}

\begin{figure}[th]
			\centering
			\includegraphics[width=3.7 in]{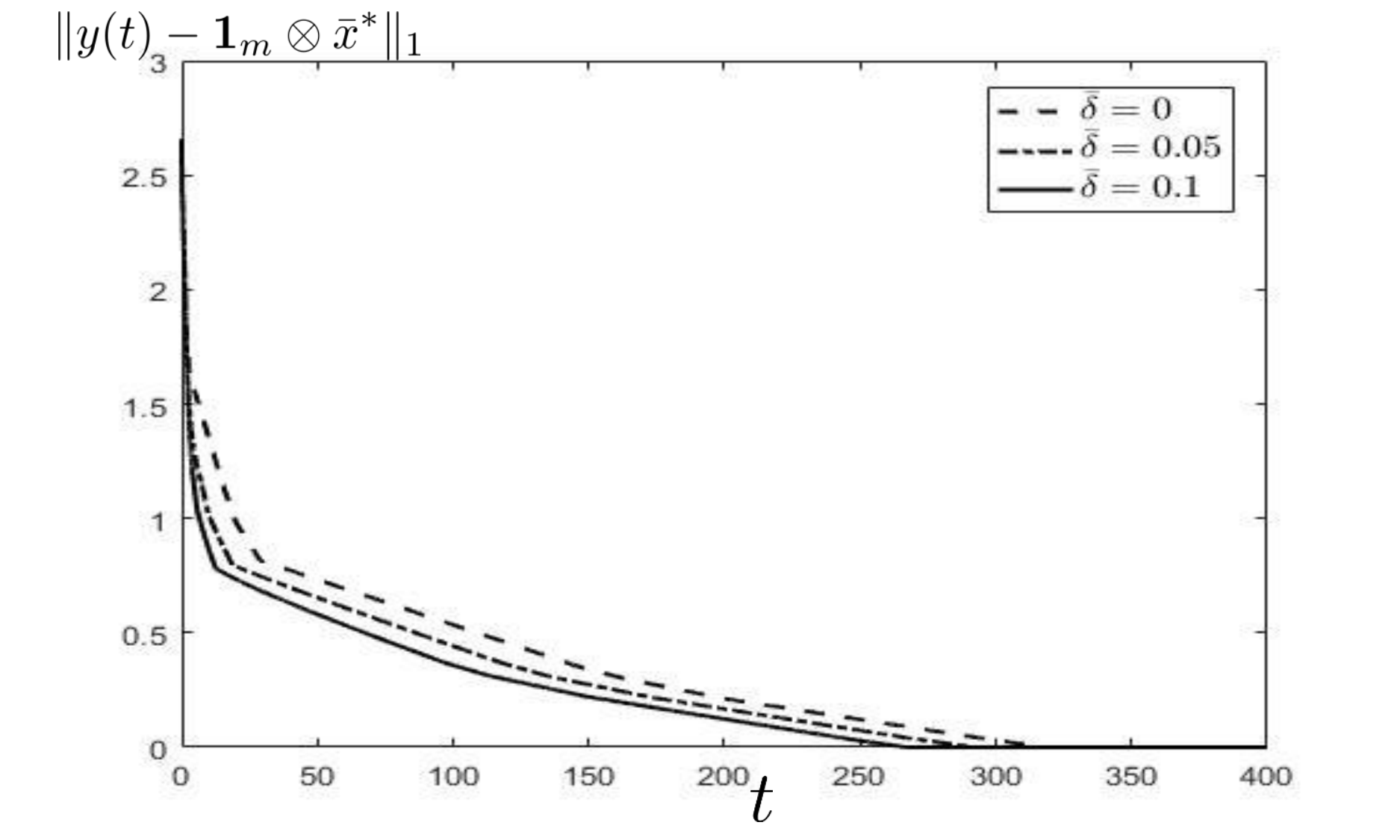}  \caption{Distributed solver for achieving minimum $l_1$ norm solution under update (\ref{l1b}) where $k(t)=\frac{\bar{\delta}}{t+1}+\delta$ with fixed $ \delta=0.01$ and different values of $\bar \delta$.}\label{f5}
		\end{figure}

\textbf{Example 3}: Finally, we utilize the distributed update (\ref{l1b}) to achieve a minimum $l_1$ solution to $Ax=b$ denoted by $\bar{x}^*$ in finite time. Here $k(t)$ is chosen to take the form $\frac{\bar \delta}{1+t}+\delta$ with $\bar \delta$ and $\delta$ constants. We still let $y=\matt{y_1'&y_2'&y_3'&y_4'}'$ where $y_i(t)$ denote the state of agent $i$ that is the estimate of agent $i$ to $\bar{x}^*$. Then $\|y(t)-\textbf{1}_m \otimes \bar{x}^*\|_1$ measures the difference between all agents' estimations and $\bar x^*$. As shown in Figure \ref{f4} and Figure \ref{f5}, all $y_i(t)$ reach the same minimum $l_1$-norm solution in finite time regardless of different choices of $\bar \delta$ and $\delta$. Moreover, by fixing $\bar{\delta}$ and increasing the value of $\delta$ in $k(t)$, one achieves a significantly faster convergence as shown in Figure \ref{f4}. Similarly, increasing $\bar \delta$ with a fixed $\delta$ also leads to a faster convergence, although not that dramatically, as shown in Figure \ref{f5}. 
	
\begin{figure}[th]
					\centering
					\includegraphics[width=3.7 in]{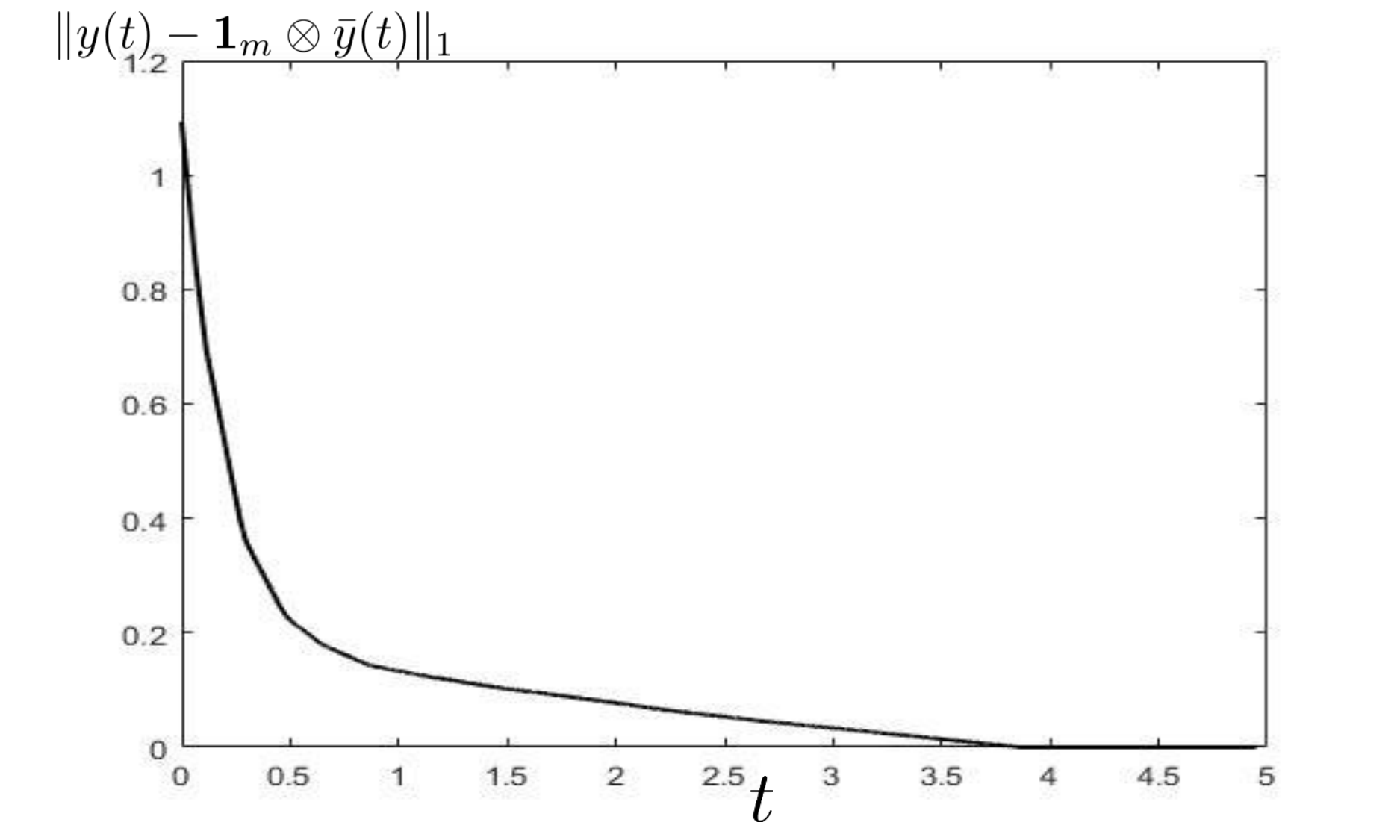}  \caption{Consensus of distributed solver under update (\ref{l1b}) with $k(t)=\frac{0.1}{t+1}+0.01$}\label{f6}
	\end{figure}

We also note from Figure \ref{f4} and Figure \ref{f5} that the convergence time required in this distributed way for minimum $l_1$-norm solutions is dramatically longer, roughly speaking, $\frac{1}{\delta}$ times longer, than that in the centralized case in Figure \ref{f3}. The major reason for this is that the centralized update appearing in the distributed update (\ref{l1b}) is scaled by $k(t)$, which is smaller than 1. The time required for consensus in this four-agent network example is minor under the distributed update (\ref{l1b}) as indicated in Figure \ref{f6}. Let $\bar y(t)$ denote the average of all four agents' states. The evolution of $\|y(t)-{\bf 1}_m\otimes \bar{y}(t)\|_1$ in Figure \ref{f6} suggests that all agents' states reach a consensus in a finite time similar to that in  Figure \ref{f2}. We anticipate that when it comes to a very large network, the convergence time for consensus might play a more significant role in convergence of the distributed update (\ref{l1b}).
\section{Conclusion} \label{Sec_con}
We have developed continuous-time distributed algorithms for achieving solutions, and minimum $l_1$-norm solutions, respectively, to linear equations $Ax=b$ in finite time. The algorithms result from combination of the projection-consensus flow proposed in \cite{GB16ACC} and the finite-time gradient flow for consensus devised in \cite{J2006Auto}, and work for fixed undirected multi-agent networks. Future work includes the generalization of
the proposed update to networks that are directed and time-varying.
\bibliographystyle{unsrt}
\bibliography{Shaoshuai,LinearEquations}

\end{document}